\documentclass[aps,prl,twocolumn,groupedaddress,amsmath,amssymb,showpacs]{revtex4}

\usepackage[dvips]{psfrag,graphicx}
\usepackage{dcolumn}
\usepackage{bm}

\begin{document}

\preprint{KUCP-0198}
\preprint{hep-th/yymmxxx}

\title{Confining Phases of a Compact $U(1)$ Gauge Theory 
\\
from the Sine-Gordon/Massive Thirring Duality}


\author{Kentaroh Yoshida}
\affiliation{Graduate School of Human and Environmental Studies,
\\ Kyoto University, Kyoto 606-8501, Japan.\\
{\tt E-mail: yoshida@phys.h.kyoto-u.ac.jp}}


\date{\today}

\begin{abstract}
We consider the phase structure of a pure compact $U(1)$ gauge
 theory in four dimensions at finite temperature by treating this system as 
a perturbative deformation of the topological model. 
Phases of a gauge theory can be investigated from the phase structure of the 
topological model. The thermal
 pressure of the topological model has been calculated, 
from which its phase structure can be derived. We have obtained phases of
 a compact $U(1)$ gauge theory. Moreover, 
the critical-line equation has been explicitly evaluated.
\end{abstract}


\maketitle



A novel scenario to study the confinement is a perturbative deformation 
of the topological model \cite{HT1, HT2,
Izawa, Kondo1, Kondo3, KY1, KY2, KY3} 
and it has clarified various properties of the confinement,
string tension, phase structure, etc. 
The phase structure of a gauge theory can be described by that of 
the topological model. In this scenario  we choose a 
modified gauge fixing which leads to an $OSp(4,2)$ symmetry in the
topological model. This symmetry enables us to apply 
the Parisi-Sourlas (PS) dimensional reduction \cite{PS} for the topological
model. Due to the PS reduction the topological model is equivalent 
to the two-dimensional non-linear sigma model (NLSM$_2$).  
In particular, the behavior of topological objects in the NLSM$_2$
decides whether a confining string between test
 particles appears or not. 
The advantage of this scenario is that the dynamics and mechanics of the
confinement are very tractable.    

In the case of a compact $U(1)$ gauge theory, the 
topological model is equivalent to 
an $O(2)$ NLSM$_2$. 
It has vortex solutions which form a two-dimensional Coulomb gas (CG). 
Phases of a compact $U(1)$ gauge theory are determined by the 
behavior of a CG. It has been shown  
that the confining phase
transition of a compact $U(1)$ gauge theory at zero temperature can be
described by the Berezinskii-Kosterlitz-Thouless (BKT) 
phase transition \cite{BKT1,BKT2,BKT3} and obtained that the
critical gauge coupling is $g_{\rm cr} = \pi$ \cite{Kondo-QED} due to 
the compactness of the gauge group as in three dimensions
\cite{Poly1,Poly2}.   
Now we consider the continuum formulation but 
the confining phase exists as in the lattice gauge theory because 
we use singular configurations with regularization. 
If we remove the regularization, then the effect of topological objects
and the confining phase disappears as well known in Guth's work \cite{Guth}.  
In this sense, regularization for monopoles is important for the
existence of the confining phase.    
It is well known that an $O(2)$ NLSM$_2$ is equivalent
to some models such as sine-Gordon (SG) model, massive Thirring (MT)
model, and XY
model. In our previous works \cite{KY1,KY2,KY3} 
we have studied phases of a pure 
compact $U(1)$ gauge theory at finite temperature using
 these equivalences and obtained the results consistent to the 
prediction by Svetitsky and Yaffe \cite{Sv-Ya,Sv}.

In this letter we will report 
the existence of the confining phase in a pure compact $U(1)$ gauge
theory in four dimensions at finite temperature as 
in the lattice gauge theory. 

Our consideration is based on demonstrating 
the phase structure of the topological model 
using the thermal pressure. We can calculate this quantity 
from the SG/MT duality and the equivalence between the 
one-dimensional CG and MT model at high temperature \cite{Steer1}. 
Moreover, we have obtained the explicit critical-line 
equation.


We start with the action of the original $U(1)$ gauge theory and 
treat the theory as a perturbative deformation of the topological
model that can be mapped to the SG model \cite{Kondo-QED,KY3}. Thus we can obtain  
the relationship between parameters in
the compact $U(1)$ gauge theory and SG model as follows, 
\begin{equation}
\label{gauge-SG}
 \lambda = \frac{128\pi^6}{g^4}\zeta\Lambda^2,~~~~m = \frac{4\pi^{3/2}}{g}\zeta^{1/2}\Lambda.
\end{equation}
Here $m$ and $\lambda$ are a mass and a coupling constant in the SG model
respectively,  and $g$ is a gauge coupling in the compact $U(1)$ gauge
theory (For notations and details, see Ref. \cite{KY3}). The
$\zeta$ is defined by $\zeta \equiv \kappa^{2\pi/g^2}$ where the
dimensionless parameter $\kappa$ is
defined by $\kappa \equiv R_0/a $ and $R_0$ 
is a radius for the regularization of vortices in an $O(2)$ NLSM$_2$. 
This quantity is 
inherently related to (regularized) Dirac monopoles in four dimensions
and important for the existence of the confining phase. The limit $\zeta
\rightarrow 0$ corresponds to remove the regularization, and the monopole
effect vanishes and the confining phase disappears. 

The $\Lambda \equiv 1/a$ is introduced in a CG through the regularization of the
 two-dimensional potential,
\begin{equation}
 V(r) \sim \ln\frac{r}{a},
\end{equation} 
and denotes a cut-off for the small distance. 
In four dimensions, $\Lambda$ corresponds 
to a cut-off for the short-range 
interaction between monopoles. Moreover,   
the string tension $\sigma_{\rm st}$ in a gauge theory (at zero
temperature) is given by
\begin{equation}
 \sigma_{\rm st} \simeq \zeta\frac{1}{a^2} = \zeta\Lambda^2. 
\end{equation}
We cannot obtain the string tension at finite temperature yet, but 
we can guess even in the finite-temperature case 
that the behavior of a CG would determine whether 
the confining string appears or not.
The $\Lambda$ plays an important role and decides the scale of the
theory (The $\Lambda$ has been missed until this paper. The expression
 in this paper is correct.).
Note that Eq.\,(\ref{gauge-SG}) does not depend on the
temperature and holds at any temperature. It has been shown that 
the duality between the SG and MT model holds 
at zero temperature \cite{Coleman} and even finite temperature \cite{Del,Steer2}
if the following relation
\begin{equation}
\label{SG-MT}
 \frac{4\pi m^2}{\lambda} = 1 + \frac{g_{\rm\scriptscriptstyle{MT}}^2}{\pi}, ~~~~\frac{m^4}{\lambda} = \rho m_{\rm\scriptscriptstyle{MT}},
\end{equation}
is satisfied. Here $m_{\rm\scriptscriptstyle MT}$ and $g_{\rm\scriptscriptstyle MT}^2$ are a renormalized mass and
 a dimensionless coupling constant, respectively. Here the
renormalization scale is set as $\rho = m_{\rm\scriptscriptstyle MT}$.  Combining
 Eq.\,(\ref{gauge-SG}) with  
Eq.\,(\ref{SG-MT}), we can write parameters of the MT model by 
those of the gauge theory as 
\begin{equation}
\label{gauge-MT}
 m_{\rm\scriptscriptstyle MT} = \sqrt{2\zeta}\Lambda,~~~g_{\rm\scriptscriptstyle MT}^2 = \frac{g^2}{2\pi} - \pi.
\end{equation}
Here 
we should remark that the physical temperature $T$ is common in the SG model,
MT model and gauge theory by derivation in our scenario.


The equivalence between 
the one-dimensional CG and MT
model at high temperature has been shown 
in the dimensional reduction (DR) regime \cite{Steer1}
\begin{eqnarray}
 T \gg m_{\rm\scriptscriptstyle MT},~~~g_{\rm\scriptscriptstyle MT}^2 > 0, \\
 T \gtrsim m_{\rm\scriptscriptstyle MT},~~~g_{\rm\scriptscriptstyle MT}^2/\pi \gg 1.
\end{eqnarray} 
A one-dimensional CG system is an exactly solvable
and has two phases, which are a molecule phase and a plasma phase 
\cite{Lenard,Ed-Le}. Thus a one-dimensional CG system undergoes a 
BKT-like phase transition at certain temperature of the CG system. 
This phase transition can be explained by 
the intensity of the 
thermal pressure, 
\begin{equation}
 P_{\rm 1CG}(z,\theta,\sigma) = 2\pi\sigma^2\gamma_0(\hat{z}),~~\hat{z}\equiv\frac{z\theta}{2\pi \sigma^2},
\end{equation}
where $z$ and $\theta$ are the fugacity and temperature of the CG
system respectively, and $\sigma$ is a charge of the particle
forming the CG. The  
$\gamma_0$ is the highest eigenvalue of Mathieu's differential
equation
\begin{equation}
 \left[\frac{d^2}{d\phi^2} + 2\hat{z}\cos\phi\right]y(\phi) = \gamma y(\phi)
\end{equation}
with $y(\phi + 2\pi) = y(\phi)$. 
The thermal pressure of the MT model 
can be written as \cite{Steer1} 
\begin{eqnarray}
\label{press}
P_{\rm\scriptscriptstyle MT}(T,m_{\rm\scriptscriptstyle MT},g_{\rm\scriptscriptstyle MT}) = \frac{\pi}{6}T^2 + P_{\rm 1CG}(T,m_{\rm\scriptscriptstyle MT},g_{\rm\scriptscriptstyle MT}), \\                
\label{thermal-p}
 P_{\rm 1CG}(T,m_{\rm\scriptscriptstyle MT},g_{\rm\scriptscriptstyle MT}) = \frac{2\pi T}{1 + g_{\rm\scriptscriptstyle MT}^2/\pi}\gamma_0 (\hat{z}),
\end{eqnarray} 
where 
$\hat{z}$ is defined by 
\begin{equation}
 \hat{z} \equiv \frac{m_{\rm\scriptscriptstyle MT}^2}{4\pi T^2}\left(1 + \frac{g_{\rm\scriptscriptstyle MT}^2}{\pi}\right)\left(\frac{2T}{m_{\rm\scriptscriptstyle MT}}\right)^{(1 + g_{\rm\scriptscriptstyle MT}^2/\pi)^{-1}}.
\end{equation}
The first term in Eq.\,(\ref{press}) is the finite-size effect and 
the dominant CG contribution of the thermal pressure comes from
$\gamma_0(\hat{z})$. Since $\gamma_0(\hat{z})$ is a monotonously
increasing function,  
$\hat{z}$ becomes the order-parameter of the one-dimensional CG and MT
model due to the behavior of $\gamma_0(\hat{z})$. If 
$\hat{z} \ll 1$, then the CG is in a molecule 
 phase and the MT model is in the chirally symmetric phase. 
If $\hat{z} \gg 1$, then the CG is in a plasma phase and the MT
model is in the chirally broken phase. 
\begin{figure}
\includegraphics[scale=.8]{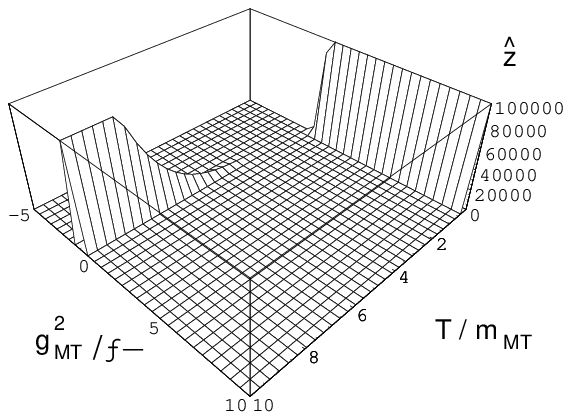}
\caption{The $\hat{z}$ is numerically plotted as a function of  $T/m_{\rm\scriptscriptstyle MT}$ and 
 $g_{\rm\scriptscriptstyle MT}^2/\pi$. A cliff and a  slope 
 exist. A cliff exists in the strong-coupling and low-temperature region. 
 There is a slope in the negative-coupling region of the MT model. 
This slope appears as another confining phase in a gauge
 theory. However, we cannot rely the result in this region as noted later. } 
\label{z-MT:fig}
\end{figure}
The numerical plot of $\hat{z}$ is shown in FIG.\,\ref{z-MT:fig}.

\begin{figure}
\includegraphics[scale=.8]{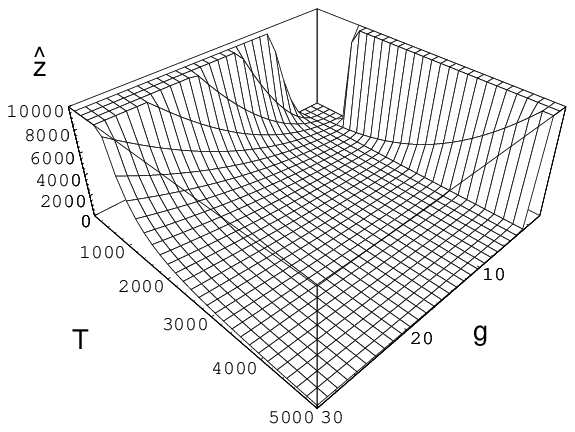}
\caption{The $\hat{z}$ is numerically plotted at
 fixed $\kappa=1$ and $Q\equiv \zeta^{1/2}\Lambda = 1000$. One
 can see two precipices. One corresponds to a traditionally known 
 confining phase but another is unknown. The weak-coupling and
 high-temperature region is out of the DR regime and the result would
 not be valid.  }
\label{plot:fig}
\end{figure}

We can translate phases in the CG and MT model as those of a gauge
theory using Eq.\,(\ref{gauge-MT}). 
 The order-parameter $\hat{z}$ can be rewritten as 
\begin{equation}
\label{order-p-gauge}
 \hat{z} = \frac{g^2}{2\pi^3}\left(\frac{2T^2}{\zeta\Lambda^2}
\right)^{\pi^2/g^2 - 1}. 
\end{equation}
The phase of a gauge theory at high temperature 
is determined by a one-dimensional
CG \cite{KY2}. Therefore, if $\hat{z} \ll 1$, 
then it is in a deconfining phase, 
and if $\hat{z} \gg 1$, then it is in a confining phase. We have
numerically plotted Eq.\,(\ref{order-p-gauge}) as shown in
FIG.\,\ref{plot:fig}. We can see two precipices which corresponds to 
confining phases. 
One corresponds to the
traditional confining phase predicted in the lattice gauge theory 
\cite{Sv-Ya,Sv}. Another is an unpredicted confining phase, 
though it is out of the region that the result is valid as noted later. 

\begin{figure}
\includegraphics[scale=.8]{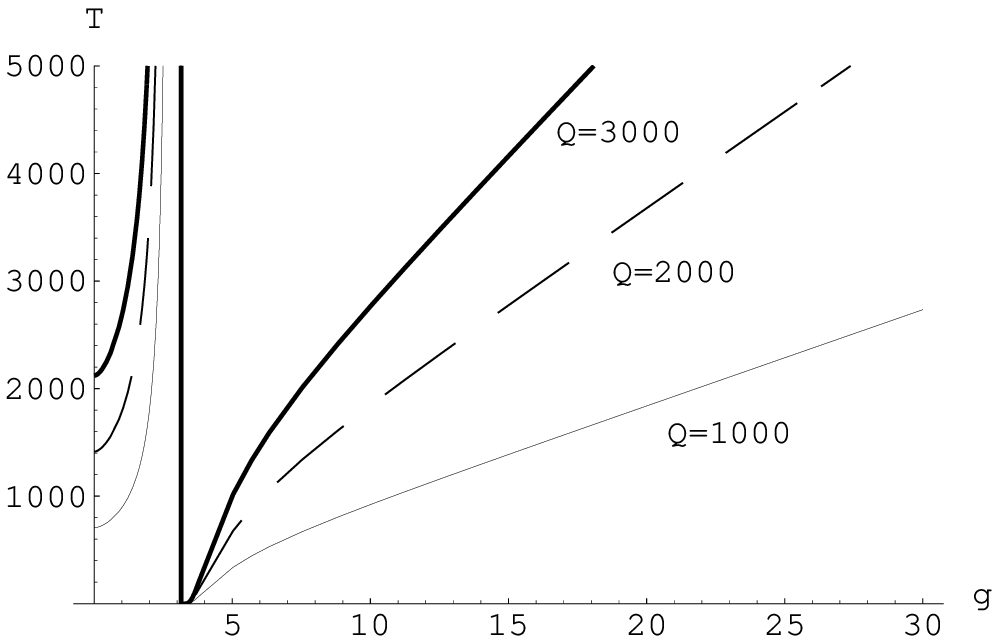}
FIG.\,\ref{cri:fig}a
\includegraphics[scale=.8]{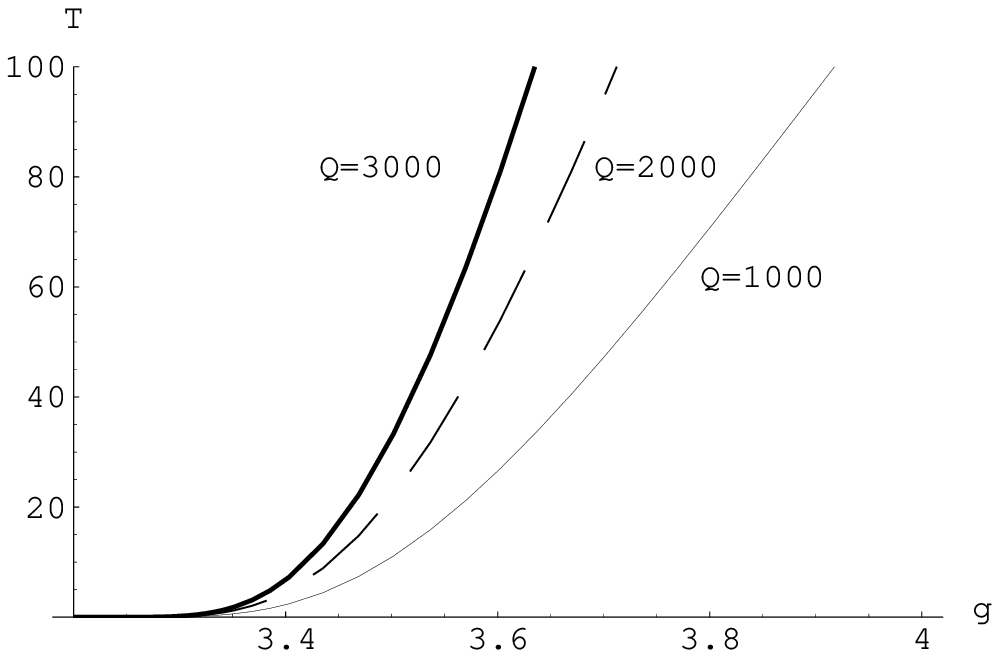}
FIG.\,\ref{cri:fig}b
\caption{The critical-line equation\,(\ref{cri-eq}) can be numerically
 evaluated. In FIG.\,\ref{cri:fig}a the behaviors of
 the  critical-line equation are shown at $\kappa =1$ and $Q\equiv \zeta^{1/2}\Lambda = 1000, 2000  $ and $3000$. 
In FIG.\,\ref{cri:fig}b those are plotted in the region near the critical coupling. }
\label{cri:fig}
\end{figure}

Moreover we can evaluate the critical-line equation by
setting $\hat{z}\simeq 1$, because $\gamma_0(\hat{z})$ increases  
very rapidly more than $\hat{z} \simeq 1$. 
As a result, we obtain
\begin{equation}
\label{cri-eq}
 T \simeq \frac{\zeta^{1/2}\Lambda}{\sqrt{2}}\left(\frac{2\pi^3}{g^2}\right)^{g^2/2(\pi^2 - g^2)}.
\end{equation}
We have also numerically plotted Eq.\,(\ref{cri-eq}) as shown in
FIG.\,\ref{cri:fig}. This result explicitly 
shows the critical behaviors of the traditional confining phase 
 and another confining phase.
In the strong-coupling region, $g^2 \gg \pi^2$, the critical-line 
equation\,(\ref{cri-eq}) becomes 
\begin{equation}
\label{line}
 T \simeq \frac{\zeta^{1/2}\Lambda}{2\pi^{3/2}}g.
\end{equation}
This equation\,(\ref{line}) is identical with the asymptotic form of the
critical-line equation  
obtained by the calculation of the one-loop effective potential in the
SG model \cite{KY3}. This coincidence surely 
confirms our results in this paper.  

Here we should comment on the validity of our results.  
Recall that we have used the equivalence between the one-dimensional CG
and MT model at high temperature in our derivation. This
equivalence is valid in the DR regime where the thermo-dynamical
limit exists. This regime in a gauge theory is expressed by 
\begin{eqnarray}
\label{DR1}
 T \gg \sqrt{2\zeta}\Lambda,~~g^2 > 2\pi^2, \\
 T \gtrsim  \sqrt{2\zeta}\Lambda,~~g^2 \gg 4\pi^2.
\label{DR2}
\end{eqnarray}
The above constraints for the temperature have no problem,   
 but those for the gauge coupling are obstacles 
to propose the existence of another confining phase. 
As discussed in Ref. \cite{Steer1}, we may formally take the gauge 
coupling $g$ arbitrary from the viewpoint of the gauge theory and the  
CG, although extra renormalizations at least
would be required from the standpoint of the MT model. 
However, it is quite well known that the density of
the magnetic monopoles decreases rapidly as the coupling
constant gets smaller \cite{BMK}. That, the monopole effect would 
almost vanish in this region. Therefore another confining 
phase would not exist.
Also, the existence of this phase depends on the value of $\kappa$. If we take
 $\kappa > 1$, then this phase disappears. 

Also, the traditional confining phase is consistent as shown in
FIG.\,\ref{cri:fig}, and 
we can also see the universal behavior
near the critical coupling $g_{\rm cr}=\pi$. 
This region is also out of the DR regime. Nevertheless,
the critical-line equation behaves as expected. The reason is unknown.

\begin{figure}
\includegraphics[scale=.6]{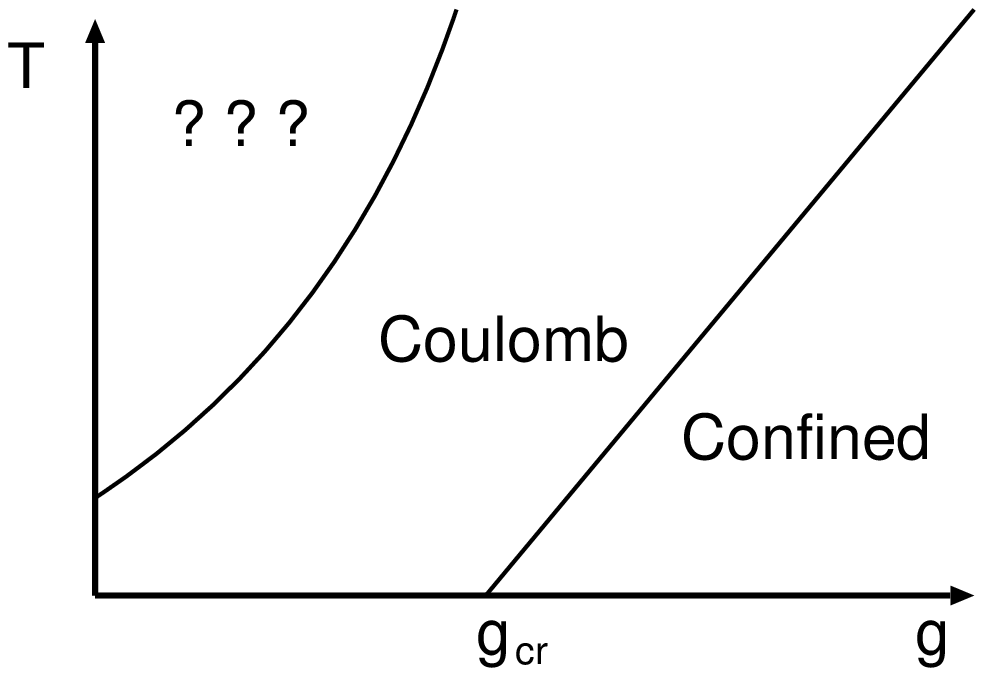}
\caption{Phase structure.}
\label{newphase:fig}
\end{figure}

In summary we have investigated the phase 
structure of a pure compact $U(1)$ gauge
theory at finite temperature using the scenario of a
perturbative deformation of the topological model.  
The topological model has been mapped to the
two-dimensional MT model through the
PS dimensional reduction and SG/MT duality. 
Due to the equivalence between the one-dimensional CG
and MT model at high temperature, we have obtained 
the thermal pressure of the topological model.
In conclusion our results suggest that the confining phase would exist 
and we propose the phase structure of a pure compact $U(1)$ gauge
theory in four dimensions at finite temperature
as shown in FIG.\,\ref{newphase:fig}. 
The confining phase at weak-coupling and high temperature region 
has not been predicted in Ref. \cite{Sv-Ya,Sv} and our results are
invalid in this region. Therefore this phase would be an error 
in our calculation.  
Moreover we have explicitly evaluated the critical-line equation. 
This result shows the traditional confining phase as known in the
lattice \cite{Sv-Ya,Sv}, and also includes the result obtained in
our previous work \cite{KY3}. 
The detailed analysis will be done in the forthcoming paper \cite{KY5}. 

The phase transition of a gauge theory could be translated to the chiral
symmetry restoration in the two-dimensional MT model at finite 
temperature.  
It is expected to have something to do with the monopole
condensation of a gauge theory in four dimensions. 
It can be also described by the behavior of a one-dimensional
CG, that is a BKT-like phase transition. Concerning with these 
descriptions, we may consider that 
topological objects in the SG model should play an important role in
the thermal phase transition of a gauge theory. 
This perspective is also attractive. 
 
In addition we can investigate the phase structure of a gauge theory 
by calculating the Gaussian effective potential (GEP) in the SG 
model and derive the critical-line equation.  
In this approach, we would be able to study the low-temperature region 
and evaluate the critical-line 
equation more precisely \cite{KY6}. We have almost calculated and 
obtained confirmed results.

\begin{acknowledgments}
The author would like to thank W. Souma for useful discussion and
valuable comments.  
He also acknowledges H. Aoyama and K. Sugiyama for
 their support in his working.
\end{acknowledgments}

\end{document}